\documentstyle{mn}

\newif\ifAMStwofonts

\ifoldfss
  \ifCUPmtlplainloaded \else
    \NewTextAlphabet{textbfit} {cmbxti10} {}
    \NewTextAlphabet{textbfss} {cmssbx10} {}
    \NewMathAlphabet{mathbfit} {cmbxti10} {} 
    \NewMathAlphabet{mathbfss} {cmssbx10} {} 
  \fi
  \ifAMStwofonts
    \ifCUPmtlplainloaded \else
      \NewSymbolFont{upmath} {eurm10}
      \NewSymbolFont{AMSa} {msam10}
      \NewMathSymbol{\upi}     {0}{upmath}{19}
      \NewMathSymbol{\umu}     {0}{upmath}{16}
      \NewMathSymbol{\upartial}{0}{upmath}{40}
      \NewMathSymbol{\leqslant}{3}{AMSa}{36}
      \NewMathSymbol{\geqslant}{3}{AMSa}{3E}

       \let\le=\leqslant
       \let\ge=\geqslant
    \fi
  \fi
\fi 

\ifnfssone
  \newmathalphabet{\mathit}
  \addtoversion{normal}{\mathit}{cmr}{m}{it}
  \addtoversion{bold}{\mathit}{cmr}{bx}{it}
  \newmathalphabet{\mathbfit} 
  \addtoversion{normal}{\mathbfit}{cmr}{bx}{it}
  \addtoversion{bold}{\mathbfit}{cmr}{bx}{it}
  \newmathalphabet{\mathbfss} 
  \addtoversion{normal}{\mathbfss}{cmss}{bx}{n}
  \addtoversion{bold}{\mathbfss}{cmss}{bx}{n}
  \ifAMStwofonts
    \ifCUPmtlplainloaded \else
      \UseAMStwoboldmath
      \makeatletter
      \new@mathgroup\upmath@group
      \define@mathgroup\mv@normal\upmath@group{eur}{m}{n}
      \define@mathgroup\mv@bold\upmath@group{eur}{b}{n}
      \edef\UPM{\hexnumber\upmath@group}
      \new@mathgroup\amsa@group
      \define@mathgroup\mv@normal\amsa@group{msa}{m}{n}
      \define@mathgroup\mv@bold\amsa@group{msa}{m}{n}
      \edef\AMSa{\hexnumber\amsa@group}
      \makeatother
      \mathchardef\upi="0\UPM19
      \mathchardef\umu="0\UPM16
      \mathchardef\upartial="0\UPM40
      \mathchardef\leqslant="3\AMSa36
      \mathchardef\geqslant="3\AMSa3E

       \let\le=\leqslant
       \let\ge=\geqslant
    \fi
  \fi
\fi 

\ifnfsstwo
  \DeclareMathAlphabet{\mathbfit}{OT1}{cmr}{bx}{it}
  \SetMathAlphabet\mathbfit{bold}{OT1}{cmr}{bx}{it}
  \DeclareMathAlphabet{\mathbfss}{OT1}{cmss}{bx}{n}
  \SetMathAlphabet\mathbfss{bold}{OT1}{cmss}{bx}{n}
  \ifAMStwofonts
    \ifCUPmtlplainloaded \else
      \DeclareSymbolFont{UPM}{U}{eur}{m}{n}
      \SetSymbolFont{UPM}{bold}{U}{eur}{b}{n}
      \DeclareSymbolFont{AMSa}{U}{msa}{m}{n}
      \DeclareMathSymbol{\upi}{0}{UPM}{"19}
      \DeclareMathSymbol{\umu}{0}{UPM}{"16}
      \DeclareMathSymbol{\upartial}{0}{UPM}{"40}
      \DeclareMathSymbol{\leqslant}{3}{AMSa}{"36}
      \DeclareMathSymbol{\geqslant}{3}{AMSa}{"3E}

       \let\le=\leqslant
       \let\ge=\geqslant
    \fi
  \fi
\fi 

\ifCUPmtlplainloaded \else
  \ifAMStwofonts \else 
    \def\upi{\pi}
    \def\umu{\mu}
    \def\upartial{\partial}
  \fi
\fi

\title[M4-18: The star and the nebula]
{M4--18: The planetary nebula and its WC10 central star}
\author[O. De Marco \& P.A. Crowther]
       {Orsola De Marco$^{1,2}$ \& Paul A. Crowther$^{1}$ \\
        $^1$Department of Physics and Astronomy, 
        University College London, 
        Gower Street, London WC1E 6BT, UK \\
        $^2$Institute f\"ur Astronomie,
        ETH Zentrum,
        Scheuchzter-Strasse 7, Z\"urich CH-8092, Switzerland}
\date{Accepted 1999 February; Received 1999 January 5}

\pagerange{\pageref{firstpage}--\pageref{lastpage}}
\pubyear{1999}
\begin{document}
\maketitle
\label{firstpage}

\begin{abstract}
We present a detailed analysis of the planetary nebula M4--18 
(G146.7+07.6) and its WC10-type Wolf-Rayet central star,
based on high quality optical spectroscopy 
(WHT/UES, INT/IDS, WIYN/DensPak) and imaging (HST/WFPC2).
{}From a non-LTE model atmosphere analysis of the stellar 
spectrum, we derive T$_{\rm eff}$=31\,kK, 
log($\dot{M}$/$M_\odot$~yr$^{-1}$)=--6.05,
$v_\infty$=160~km~s$^{-1}$ and abundance number ratios of
H/He$<$0.5, C/He=0.60 and O/He=0.10. These parameters are 
remarkably similar to He~2--113 ([WC10]). Assuming 
an identical stellar mass to that determined by De Marco et al.
for He~2--113, we obtain a distance of 6.8\,kpc to M4--18
($E_{\rm B-V}$=0.55\,mag from nebular and stellar techniques).
This implies that the planetary nebula of M4--18 has a dynamical 
age of $\sim$3\,100 years, in contrast to $\ge$270 years for 
He~2--113. This is supported by the much higher electron density of
the latter. These observations may only be reconciled with evolutionary 
predictions if  [WC]-type stars exhibit a range in stellar masses. 

Photo-ionization modelling of M4--18 is carried out using our stellar
WR flux distribution, together with blackbody and Kurucz energy
distributions obtained from Zanstra analyses. We conclude that the ionizing 
energy distribution from the Wolf-Rayet model provides the best consistency 
with the observed nebular properties, although discrepancies remain. 
\end{abstract}

\begin{keywords}
stars: individual: M4--18; stars: Wolf-Rayet; planetary nebulae: general. 
\end{keywords}
\vfill
\newpage

\section{Introduction}\label{sc:intro}

Low excitation Wolf--Rayet (WR) central stars of planetary nebula (PN; 
denoted [WR] following van der Hucht et al. 1981)\nocite{HUC+81}
are thought to represent the beginning of hydrogen--deficient central star
evolution, just following the ejection of the PN which occurs at the 
top of the asymptotic giant branch (AGB). 
However, discrepancies between observed abundances and those predicted by 
theory, led to
the proposal that H-deficient central stars of PN (CSPN)
are the result of the re-birth of a white dwarf, after
a late helium--shell pulse (Iben et al. 1983)\nocite{IKTR83}. 
Following this interpretation, the search began for characteristics
common to planetary nebulae 
with WC nuclei, that could be used to distinguish them
from those with H-rich central stars, and thus establish that the two 
classes have followed different evolutions.
However, it soon emerged that nebulae associated with [WC] central stars are 
indistinguishable from those around H-rich CSPN (Gorny \& Stasinska 1995)
\nocite{GS95}.

To seek a solution to this problem and determine the evolution of WR type 
nuclei, a flurry of empirical and modelling analyses of WC central stars were
carried out by several groups (e.g. Leuenhagen, Hamann \& Jeffery 1996,
hereafter LHJ)\nocite{LHJ96}.
The present study focuses on M4--18 (PN G146.7+07.6, IRAS 04215+6000), 
a PN with a [WC10] central star
(following the classification scheme of Crowther, De Marco \& Barlow 1998).
\nocite{CDB98}
One may reasonably ask why is yet another study of a late 
WC-type CSPN necessary? Recent stellar (LHJ)
and nebular (Surendiranath \& Rao 1995, hereafter SR) 
models for M4--18 exist in the
literature, together with empirical determinations of its PN properties
(e.g. Goodrich \& Dahari 1985).
The justification is that previous studies considered the star 
and nebula of M4--18 in isolation, making it impossible to relate
nebular and stellar results; additionally a comparison between the 
nebular properties of M4--18 with other PN associated with 
spectroscopically similar stars 
suggests that [WCL] stars can follow different evolutionary paths; 
this key aspect has never been given due attention\footnote{Pottasch
(1996) remarked that the PNe around M4--18 and He~2--113 (another [WC10] star)
are very different, hinting that the two central
stars cannot have followed the same evolutionary path, but nobody
took this argument past Table~10 of his contribution.}. 
De Marco, Barlow \& Storey (1997) and De Marco \& Crowther (1998,
hereafter DC)
carried out a rigorous analysis of the central stars
and PNe of the [WC10] central stars
CPD--56$^{\circ}$8032 and He~2--113. We follow these methods for M4--18, 
so that the relative results can be compared with confidence.

Our analysis of M4--18 also extends the photo-ionization modelling
of CPD--56$^{\circ}$8032 and He~2--113 by DC
using flux distributions appropriate
for Wolf-Rayet stars. DC identified a major
discrepancy between the observed and predicted nebular 
properties for these PNe; either the 
lack of heavy element line blanketing in the wind 
models was to blame, or the geometry and high nebular 
densities of those PNe meant that they represented poor probes 
of the Lyman continuum flux of their central stars. As we shall
demonstrate, the PN of M4--18, with a significantly lower electron
density, provides a more rigorous test of the theoretical flux distribution.

In Section~\ref{sc:obs} we will describe the observations, while in 
Section~\ref{sc:mag} we will discuss basic observational quantities,
including reddening. Section~\ref{distance} discusses the distance towards
M4--18, while a quantitative analysis is carried out in
Section~\ref{stellar}. Archive Hubble Space Telescope (HST)
images are presented in Section~\ref{sc:hst}, while 
we carry out a nebular abundance analysis in Section~\ref{sc:neb_abu}.
In Section~\ref{sc:neb_mod} photo-ionization modelling of the PN
is carried out. Finally, we draw our conclusions in Section~\ref{sc:disc}.

\section{Observations and data reduction}\label{sc:obs}

We have obtained high quality optical spectroscopy of
M4--18 using the 4.2 m William Hershell Telescope (WHT),
the 2.5~m Isaac Newton Telescope (INT) and the 3.5~m 
Wisconsin-Indiana-Yale-NOAO telescope (WIYN). Archival International
Ultraviolet Explorer (IUE) ultraviolet (UV) spectroscopy
and HST optical imaging of M4--18 were 
obtained from the Uniform Low-Dispersion Archive (Talavera 1988)
at the Rutherford Appleton Laboratory, and the HST data 
archive at  the Space Telescope Science Institute, respectively.
The HST Wide Field and Planetary Camera 2 (WFPC2) data set 
is discussed further in Section~\ref{sc:hst}.

\subsection{Optical spectroscopy}

M4--18 was observed at the INT,  using the Intermediate Dispersion Spectrograph
(IDS), together with the 235-mm camera and a 1024$\times$1024 pixel Tektronix
CCD, between  July 17 and 23, 1996. Four settings with the 1200B/Y gratings 
provided complete wavelength coverage in the range 3800--6800\AA, at a spectral
resolution of 1.5\AA\ using a 1.5 arcsec slit. Additional observations with the
300V grating and an 8 arcsec slit (including the entire nebula) provided an
absolute flux calibration  at a spectral resolution of 9.5\AA.  The data were
reduced in a standard manner using the  {\sc iraf} package, with wavelength
calibrations achieved using  Cu-Ne and Cu-Ar arc lamps and absolute flux
calibrated data obtained by comparing our wide slit observations of M4--18 with
observations of the Oke (1990)\nocite{O90} 
standards B2~{\sc iv} star BD+33$^{\circ}$2642
and the Op star BD+28$^{\circ}$4211.

High spectral resolution ($R\sim$30\,000) observations of
M4--18 were obtained at the WHT using the
Utrecht Echelle Spectrograph (UES) in service mode on 
November 27, 1996. A 1024$\times$1024 pixel Tektronix~CCD provided
a complete spectral range of 4200--5850\AA; two exposures each of
duration 2400 sec achieved a continuum signal--to--noise ratio of 40.
The data were reduced in a standard manner using the 
{\sc iraf} package, with wavelength calibration achieved relative
to a comparison Th-Ar arc. For flux calibration, the Oke (1990) 
standard G191B2b (DA0) was used. 
Subsequent data reduction 
was carried out with the {\sc dipso} package (Howarth \& Murray 1991)\nocite
{HM91}. 

Additional high resolution (R$\sim$19\,000) spectroscopy was 
obtained on November 13, 1998 by
D. Sawyer with the WIYN 3.5~m telescope and
the DensPak fiber array, in the range 5770-6010\AA. Three
exposures each lasting 600~s were obtained with 
a signal--to--noise ratio of 20. The data were
reduced with standard {\sc iraf} routines. Wavelength calibration was with 
respect to a Th-Ar lamp. These spectra allowed us to measure the radial 
velocity shifts of the interstellar Na~{\sc i} D lines.

Finally, our high resolution INT and WHT observations were scaled to
the continuum level of the wide--slit INT spectra, which were obtained
during photometric conditions. By convolving our observed spectrophotometry  
with suitable synthetic filters (courtesy of J.R. Deacon) we obtained 
measurements of both wide-band Johnson photometry (V=14.11 and B=14.24\,mag)
and narrow-band Smith (1968) photometry ($v$=14.14 and $b$=14.16\,mag).
\nocite{S68a}
Our measurements are in reasonable agreement with Shaw \& Kaler (1985) who
\nocite{SK85}
obtained a (nebula corrected) visual brightness of $\sim$14.0\,mag.

\subsection{Ultraviolet spectroscopy}

We utilised four (SWP and LWR/LWP) low resolution IUE 
large aperture spectroscopic observations of M4--18 obtained 
between August 1980--August 1991
(LWR8401 was not used because it is affected by a cosmic 
ray hit at the wavelength of  the C~{\sc ii}] line $\lambda$2326,
Goodrich \& Dahari 1985). 
Although of poor quality, these provided constraints on the 
interstellar reddening towards M4--18, and allowed us 
to obtain estimates of the nebular carbon abundance.

\section{Basic observational quantities}\label{sc:mag}

\subsection{Radial and nebular expansion velocity}

The radial velocity was measured from Gaussian fits to the
nebular Balmer lines observed in our optical datasets.
WHT--UES observations of H$\beta$--$\gamma$ indicated a
mean heliocentric radial velocity of 
--51.6$\pm$1.0\,km~s$^{-1}$ (corresponding to an 
LSR radial velocity of --50.8$\pm$1.0\,km~s$^{-1}$).
These results are supported by our lower resolution
INT observations, which indicate a heliocentric radial 
velocity of --46.1$\pm$5.0\,km~s$^{-1}$.

\begin{figure}
\vspace{8.6cm}
\includegraphics{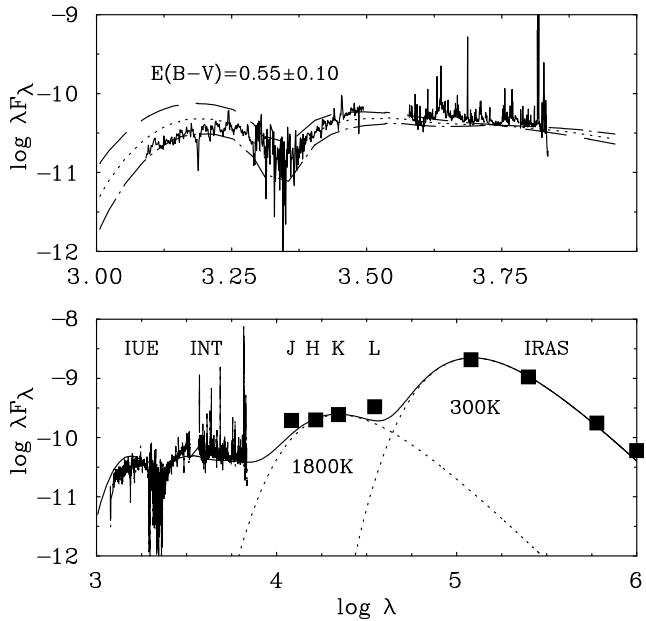}
\caption{(Upper panel): observed ultraviolet and optical 
spectrophotometry of 
M4--18 (units are erg\,cm$^{-2}$\,s$^{-1}$) from IUE and INT, together with 
theoretical energy distributions reddened by
E$_{\rm B-V}$=0.45 (dashed), 0.55 (dotted), and 0.65\,mag (dot-dashed)
using a standard Galactic extinction curve (Seaton 1979; Howarth 1983).
(Lower panel): As above, except including near and mid-IR photometry 
from SR and IRAS (filled-in squares), 
plus a two-component blackbody fit (dotted lines) to the IR 
excess caused by dust in the PN.}
\label{fg:ste_flu}
\end{figure}
\nocite{SEA79}

We used the FWHM of nebular profiles (H$\beta$, H$\gamma$ and $\lambda$5755
[N~{\sc ii}]) in our WHT--UES observations (FWHM $\sim$6 km\,s$^{-1}$) 
to determine the 
expansion velocity  of the PN of M4--18, which was revealed to be
$v_{\rm exp}$=19$\pm$0.5\,km~s$^{-1}$. [The value reported on 
the ESO PN catalogue
(Acker et al. 1992) of 12~km~s$^{-1}$ derives from an average of [O~{\sc iii}] 
(7.5~km~s$^{-1}$) and [N~{\sc ii}] (17.0~km~s$^{-1}$); we note that 
it is inappropriate to average values obtained from lines belonging 
to a mixture of ionization stages, since high ionization lines 
may originate from inner, slower parts of the PN.]

\subsection{Interstellar reddening}

Reddening determinations towards M4--18 from the literature range
from $E_{\rm B-V}$=0.48 (SR) to $E_{\rm B-V}$=0.90
(Goodrich \& Dahari 1985). We derive interstellar reddenings 
using observed Balmer line fluxes, which are 
listed in Table~\ref{tb:hflux}. Our wide-slit INT H$\beta$ nebular flux
is in reasonable agreement with the determinations of 
1.04$\times$10$^{-12}$ erg\,cm$^{-2}$\,s$^{-1}$ by Goodrich \& 
Dahari (1985), and 1.17$\times$10$^{-12}$ erg\,cm$^{-2}$\,s$^{-1}$ 
by Carrasco, Serrano \& Costero (1983, 1984)\nocite{CSC83}\nocite{CSC84}.

We adopt 
N$_e$=10$^4$ cm$^{-3}$ and T$_e$=10$^4$ K (these are 
revised slightly in Section~\ref{sc:neb_abu}), and the Galactic extinction law
of Howarth (1983)\nocite{H83} plus hydrogen recombination coefficients 
from Storey \& Hummer (1995)\nocite{SH95}. Unfortunately H$\alpha$ 
is severely blended with [N\,{\sc ii}] in our wide slit INT observations, so 
we have to rely on narrow-slit observations which imply
$E_{\rm B-V}$=0.45 mag from H$\alpha$--H$\beta$ (WHT H$\beta$-H$\gamma$
observations imply $E_{\rm B-V}$=0.61).
 
\begin{table}
\caption{Balmer fluxes (ergs~cm$^{-2}$~s$^{-1}$) 
measured for the nebular hydrogen lines of M4--18.} 
\begin{tabular}{cccc}
\hline
Line & WHT--UES  & INT--IDS &  INT--IDS \\ 
     & narrow slit & narrow slit & wide slit \\
\hline
H$\alpha$ & --                    & 2.69$\times$10$^{-12}$& --\\
H$\beta$  & 6.61$\times$10$^{-13}$& 5.75$\times$10$^{-13}$& 8.95$\times$10$^{-13}$\\
H$\gamma$ & 2.30$\times$10$^{-13}$& --                    & --\\
\hline
\end{tabular}
\label{tb:hflux}
\end{table}

We can also derive a reddening towards M4--18 following 
Milne \& Aller (1975)\nocite{MA75}, using our observed
H$\beta$ flux, together with the observed 5GHz flux of 22 mJy from
Aaquist \& Kwok (1990)\nocite{AK90}. This method implies a
higher reddening of $E_{\rm B-V}$=0.65 mag.

Therefore, extinctions in the range $E_{\rm B-V}$=0.45--0.65~mag are
implied from the {\it nebula}, so we can now test which extinction 
reproduces the observed UV and optical {\it stellar} flux distribution, 
making use of our  theoretical energy distribution from Section~\ref{stellar}. 
In Fig.~\ref{fg:ste_flu}, we compare observed UV and optical 
spectrophotometry of M4--18 with our theoretical model for interstellar extinctions
 of $E_{\rm B-V}$=0.45, 0.55 and 0.65\,mag. We find that 0.55$\pm$0.05\,mag, 
provides the best match to observations. (The only previous reddening 
determination 
towards M4--18 from spectrophotometry was by LHJ
who obtained $E_{\rm B-V}$=0.70 from a continuum fit to exclusively UV data.) 
We therefore adopt $E_{\rm B-V}$=0.55 as appropriate to both 
the nebular and stellar observations of M4--18. 

Fig.~\ref{fg:ste_flu} also includes blackbody fits to the infra-red (IR) 
excess of M4--18,
representing warm dust from its PN, based on near-IR observations from 
SR, plus colour corrected 12--100$\mu$m photometry 
from IRAS. 
We find that a two component blackbody fit is necessary, indicating the
presence of hot ($\sim$1800K) and warm ($\sim$300K) dust. DC
found somewhat cooler dust properties for CPD--56$^{\circ}$8032 and 
He~2--113.

\section{The distance to M4--18}\label{distance}

Distance estimates to M4--18 range from 1--7\,kpc
(SR; Cahn, Kaler \& Stanghellini 1991).\nocite{CKS91} 
De Marco et al. (1997)\nocite{DBS97} obtained estimates of the distances
to CPD--56$^{\circ}$8032 and He~2-113 based on the LSR radial velocities
of the interstellar Na~{\sc i} D lines, or following the 
assumption that the entire
bolometric luminosity is re-radiated in the IR by cool circumstellar dust.
The electron density of the PN of M4--18 is more than a factor of ten 
lower than that of 
CPD--56$^{\circ}$8032 and He~2-113 (Section~\ref{sc:neb_abu}), 
with IRAS fluxes about 30 times lower, suggesting that the stellar luminosity 
is unlikely to be entirely re-radiated in the infrared.

\begin{figure}
\vspace{8.8cm}
\includegraphics{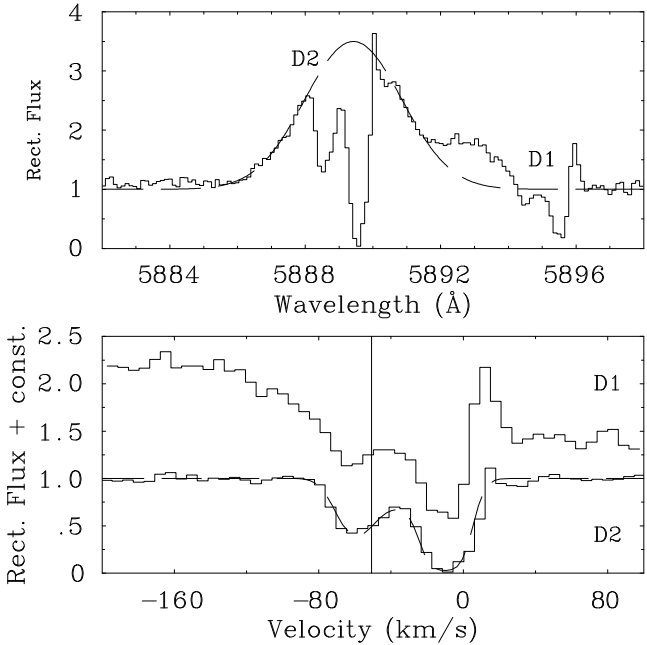}
\caption{{\bf Upper panel}: The Na~{\sc i} D lines in the WIYN spectrum
of M4--18. The dashed line shows a one-Gaussian 
fit to the main stellar emission feature.
{\bf Lower panel}: The rectified D2 spectrum in the LSR rest frame,
fitted by a three-cloud model (dashed
line). D1 is also shown. The vertical line marks the LSR radial velocity of the PN.}
\label{fig:nad}
\end{figure}

In Fig.~\ref{fig:nad} we present the WIYN/DensPak observations in the
spectral region 5880--5900\AA, showing the Na~{\sc i} D lines superimposed 
on a stellar emission blend (C~{\sc ii} M5, C~{\sc iii} M20 and 
He~{\sc ii} $\lambda$5897, De Marco et al., 1997). 
To determine the radial velocities of the individual Na~{\sc i} D
line components, we used a model that calculates 
absorption profiles for a variety of interstellar clouds with 
Gaussian line-of-sight velocity distributions (implemented into the
{\sc iscalc} routine  in the {\sc dipso} package). 
To rectify the Na~{\sc i} D lines we `snipped' the D2 absorption and
neglected the spectrum longward of $\sim$5892\AA, thereby approximating
the broad emission feature with a single Gaussian
(dashed line in the upper panel of 
Fig.~\ref{fig:nad}). 

The three-cloud component fit to the Na~{\sc i} D2 line 
is presented in Fig.~\ref{fig:nad} (lower panel), with the 
corresponding parameters listed in Table~\ref{tb:nad}. 
Although we were unable to carry out the same procedure for D1, 
it is clear from Fig.~\ref{fig:nad} that the positions of the D1
absorption line components match those of the D2 components. 

\begin{table}
\caption{Parameters of the three-cloud model shown in Fig.\ref{fig:nad}. The
velocity dispersion parameter $b$ is defined in Howarth \& Phillips (1986), 
while $N$ is Na~{\sc i} the column density. The radial velocity
(RV) components are in the LSR rest frame, 
while $D$ is the distance implied by the Brand \& Blitz (1993) galactic
rotation curve for material near the galactic plane.}
\label{tb:nad}
\begin{tabular}{ccccc}
\hline
   Cloud no. &  $b$&  Log($N$/cm$^{-2}$)&RV        &  $D$ \\
             & (km~s$^{-1}$)&&  (km~s$^{-1}$)  & (kpc) \\         
\hline
       1     &     13.0   &  13.0 & --60.0$\pm$5.0 & -- \\
       2     &     20.0   &  12.8 & --37.0$\pm$5.0 & $>$1.5\\
       3     &     12.0   &  13.6 & --10.0$\pm$5.0 & $>$1.0\\
\hline
\end{tabular}
\end{table}
\nocite{BB93} 
\nocite{HP86}

If the --60~km~s$^{-1}$ radial velocity component 
is of interstellar origin, the galactic radial velocity map of 
Brand \& Blitz (1993) indicates that M4--18 lies beyond 10~kpc. However, 
at this distance, M4--18 would lie 1.2~kpc above the galactic plane, 
where neutral sodium is likely to be scarce
(the scale height of neutral hydrogen was determined to be 144$\pm$80~pc
by Shull \& Van Steenberg 1985)\nocite{SV85},
Moreover, the Brand \& Blitz (1993) radial velocity curve was 
obtained  from measurements of H~{\sc ii} regions with a scale height of 
only 67~pc. In the direction of M4--18, it 
can be considered reasonably accurate up to a distance from the galactic plane 
of a few times this value ($\approx$200~pc; several measurements
of H~{\sc ii} regions at $\sim$400~pc above the galactic plane 
exist for this line-of-sight). Interstellar material in the direction of
M4--18, therefore extends to a distance of $\sim$1.5~kpc, corresponding to a 
radial velocity of --30~km~s$^{-1}$. 
Na~{\sc i} D line components with radial velocity more negative than this 
value, cannot be interstellar.
Considering the uncertainty on the Galactic rotation curve, 
we identify the observed --37~km~s$^{-1}$ 
component to be of interstellar origin. Consequently,
Na~{\sc i} D lines simply argue for the distance towards
M4-18 being greater than 1.5~kpc.

We attribute the --60~km~s$^{-1}$ component to 
nebular material. Although we may expect M4--18 not to have a 
neutral envelope since it is optically thin (see Section~\ref{sc:neb_mod}),
Na~{\sc i} was observed in other PNe by Dinerstein, Sneden \& Uglum (1995),
some of which are also optically thin. If Na~{\sc i} is present in the 
PN shell, its expansion velocity can be obtained from the difference
between the LSR radial velocity of the PN (--50.8~km~s$^{-1}$, 
Section~\ref{sc:mag}) and the radial velocity shift of the Na~{\sc i}
D line component at --60~km~s$^{-1}$. This results in 9~km~s$^{-1}$, 
significantly lower than the ionized shell expansion velocity 
(19~km~s$^{-1}$). Neutral shells moving considerably slower than 
the ionized envelopes were detected by Dinerstein et al. (1995)
in two other PNe (SwSt~1 and IC3568).

Therefore, we follow the alternative approach of DC
for CPD--56$^{\circ}$8032 and He~2-113 based on assuming 
a core mass of 0.62\,$M_{\odot}$ (obtained by X.-W. Liu [priv. comm.] for 
five LMC WR CSPN), and applying the helium burning post-AGB evolutionary
tracks of Vassiliadis \& Wood (1994)\nocite{VW94}
to obtain a stellar luminosity of 5250\,$L_{\odot}$. Utilising the observed
$v$-band magnitude for M4--18 from our spectrophotometry, together with the 
interstellar extinction of $E_{\rm B-V}$=0.55$\pm$0.05\,mag and bolometric
correction from our WR model (Section~\ref{ssc:wr_mod}) we obtain a distance of
$\approx$6.8\,kpc. This is in excellent agreement with recent statistical 
distances for M4--18, which range from 6.7--7.1\,kpc, as summarised by 
Zhang (1995)\nocite{Zha95} and with the distance implied by the --37~km~s$^{-1}$ Na~{\sc i} D line
component. 

\begin{figure*}
\vspace{10.8cm}
\includegraphics{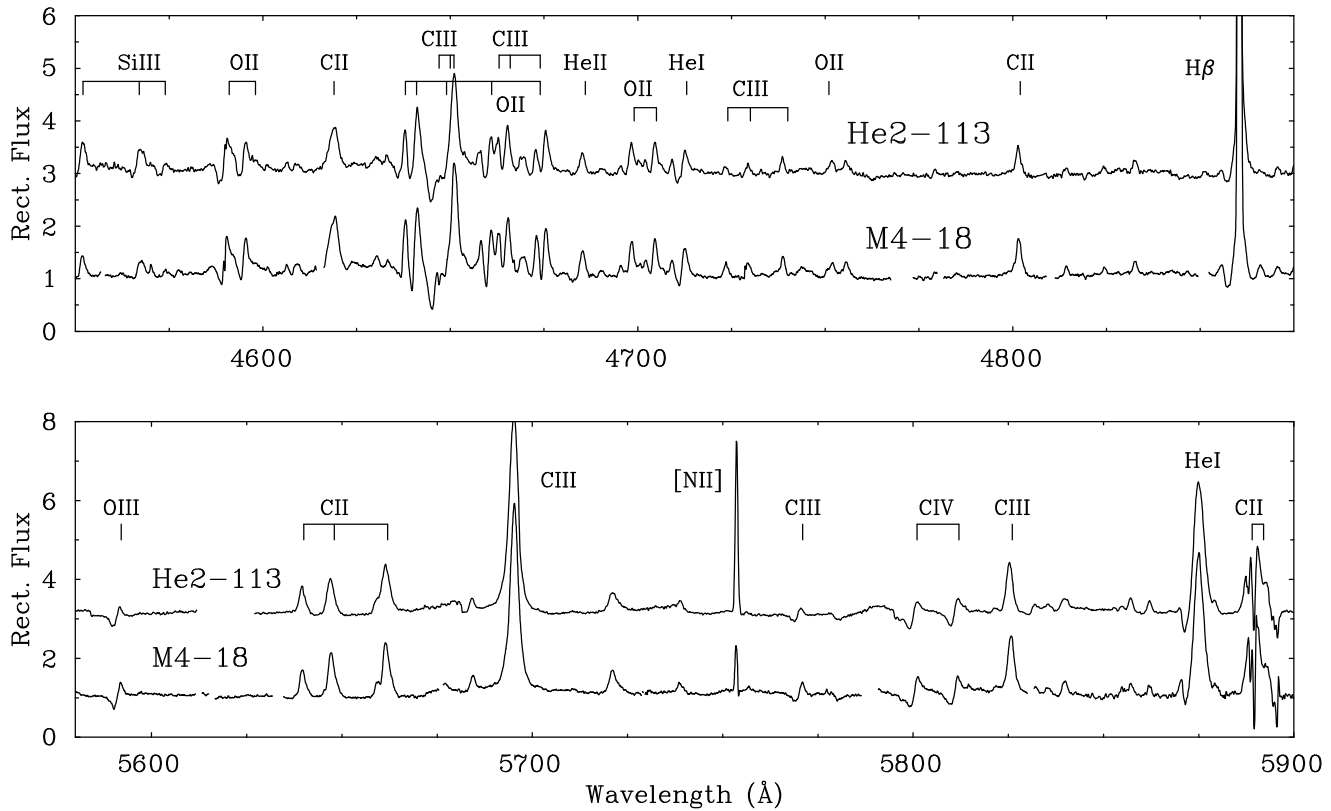}
\caption{Comparison between rectified spectra of M4--18 (WHT--UES,
WIYN--DensPak for $\lambda$$>$5845\AA)
and He~2--113 (DC), demonstrating the striking 
similarity of their appearances.}
\label{m418_hen_comp}
\end{figure*}
\nocite{DC98}

Of course, the masses of all WC central stars need not be identical. 
{}From Vassiliadis \& Wood (1994) and Bl\"{o}cker (1995), the possible range 
in luminosity for a CSPN is $\sim$2\,500--16\,000~L$_\odot$,
corresponding to a possible range in distance of 4.8--12.0\,kpc. 
Consequently, all distances lower than this minimum are excluded.
For example, SR adopted a distance 
of 1 kpc, based on a relationship between the dust temperature 
and the PN radius (Pottasch et al. 1984)\nocite{PBBEHHHJM84}.
Using our bolometric correction, an unrealistically low stellar
luminosity of 110$L_{\odot}$ would be implied 
for this distance. (Alternatively,
an unrealistically high bolometric correction of $-$6 mag would be required 
assuming the luminosity from SR namely 2\,500$L_{\odot}$).

In summary, we adopt a distance of 6.8\,kpc to M4--18, assuming a mass and 
luminosity identical to that obtained by DC for 
He~2--113. 
However, although M4--18 and He~2--113 are spectroscopically very similar 
(Section~\ref{stellar}), this assumption could be in error. The existence
of massive WC stars and CSPN with very similar spectra, such as He~2--99 
([WC9]) and HD\,164270 (WC9) (see Mendez et al. 1991)\nocite{MHMK91}, 
is a clear indication
that spectral similarities do not necessarily indicate similarities in mass.
This estimate, on the other hand 
is in agreement with the limits imposed by stellar evolution.

\section{Stellar Analysis}\label{stellar}

In this Section we will present our stellar spectrum of M4--18,
addressing the important question of whether stellar hydrogen is present.
We will also obtain stellar parameters using a sophisticated stellar atmosphere 
model appropriate for Wolf-Rayet stars.

\subsection{The stellar spectrum -- is hydrogen present?}\label{ssc:spec}

The stellar spectrum of M4--18 is dominated by emission lines
of helium (He~{\sc i}--{\sc ii}), carbon (C~{\sc ii}--{\sc iv})
and oxygen (O~{\sc ii}--{\sc iii}). Our high resolution spectroscopy
of M4--18 is presented in Fig.~\ref{m418_hen_comp}
along with a spectrum of the [WC10] central
star He~2--113 (from De Marco et al. 1997). 
It is apparent that their spectral morphologies are remarkably similar,
including line widths, suggesting comparable outflow velocities.
The nebular lines of the M4--18 PN are much stronger than in He~2--113.

\begin{figure}
\vspace{13.8cm}
\includegraphics{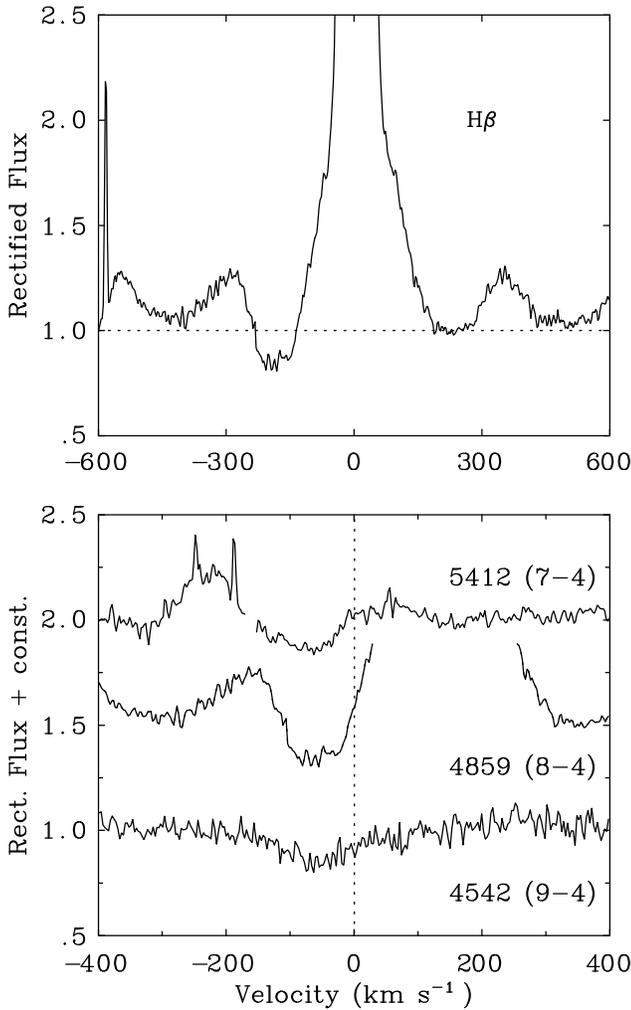}
\caption{(upper panel) Base of the rectified H$\beta$ profile in the 
WHT--UES dataset of M4--18, revealing the presence of (i) broad line
wings and (ii) a P~Cygni component; (lower panel) Comparison of
rectified He\,{\sc ii} Pickering series profiles, indicating that
the H$\beta$ P~Cygni component is due to He\,{\sc ii} $\lambda$4859
(8--4).}
\label{balmer}
\end{figure}

The upper panel of Fig.~\ref{balmer} shows the high resolution 
rectified UES spectrum, in the rest frame of the H$\beta$ line.
This feature appears to broaden at its base; additionally, 
a blue--shifted absorption (whose minimum intensity is registered
at 185 km~s$^{-1}$ to the blue of the
line's rest wavelength), is apparent from the rectified
spectrum in Fig.~\ref{balmer} (upper panel), suggesting that
stellar hydrogen is present in the wind. LHJ determined a 
H/He ratio of $<$10, 0.64 and $<$0.5
by number, for the [WC10] central stars M4--18, He~2--113 
and CPD--56$^{\circ}$8032, respectively. De Marco et al. (1997) noticed 
the same  behaviour in the Balmer lines of CPD--56$^{\circ}$8032 and 
He~2--113. However, by comparing these profiles with nebular [O~{\sc i}] 
lines at 6300 and 6363\AA, they demonstrated that the origin of 
the broad pedestal under the H$\alpha$ and H$\beta$ profiles is more 
likely to be due to an irregular nebular geometry. 

For M4--18, nebular [N\,{\sc ii}] $\lambda$5755 and H$\gamma$ 
are also present in the high resolution UES spectrum,
although these are too weak for a comparison to be made.
However, we are able to comment on the origin of the P~Cygni profile, 
which cannot be produced by peculiar nebular material.
The lower panel of Fig.~\ref{balmer} compares the P~Cygni absorption feature 
in the rest frame of He~{\sc ii} 
4859.18\AA\ (transition 8--4), together with the adjacent 
Pickering members $\lambda$4541.46 (9--4) and $\lambda$5411.37 (7--4). 
We find that He\,{\sc ii} is responsible for this feature.

To summarise, we suspect that H$\beta$ emission wings are solely 
due to high velocity components in the {\it nebular} spectrum, 
as is the case for the other two [WC10] central stars
He~2--113 and CPD--56$^{\circ}$8032.
(De Marco et al. 1997; Sahai, Wotten \& Clegg 1993). To verify this, 
one would need to observe other nebular lines at high spectral resolution.

\subsection{Spectroscopic analysis}\label{ssc:wr_mod}

Hillier (1987, 1990)\nocite{HIL90a}\nocite{HIL87a} 
theoretical model atmospheres are used in our
spectroscopic analysis.  These calculations employ an iterative
technique to solve the transfer equation in the  co-moving frame subject
to statistical and radiative equilibrium in an expanding,
spherically-symmetric, homogeneous and steady-state atmosphere.  
The stellar radius ($R_{\ast}$) is defined as the inner boundary of the model
atmosphere and is located at a Rosseland optical depth of 20. The 
temperature stratification is determined from the assumption of 
radiative equilibrium with the temperature parameter ($T_{\ast}$) 
defined by the usual Stefan-Boltzmann relation. Similarly, the effective
temperature ($T_{\rm eff}$) relates to the radius ($R_{\rm 2/3}$)
at which the Rosseland optical depth equals 2/3. 
The spectral synthesis proceeds by fitting line profiles to
observed diagnostic He\,{\sc i-ii}, C\,{\sc ii-iv}, 
O\,{\sc ii-iii} lines, together with the absolute visual magnitude. 
In the absence of high resolution ultraviolet observations 
we measured a stellar wind terminal velocity from optical
He\,{\sc i} P~Cygni profiles, resulting in 160$\pm$15\,km\,s$^{-1}$.

\begin{table*}
\caption{Derived stellar parameters of M4--18, including a comparison  with 
Leuenhagen et al. (1996, LHJ) scaled to our assumed luminosity, and 
demonstrating its very close similarity with He\,2--113 (from De Marco 
\& Crowther 1998, DC). We include the predicted H\,{\sc i}  and 
He\,{\sc i} continuum ionizing fluxes ($Q_{0}$, $Q_{1}$).}
\label{tb:wr_mod}
\begin{center}
\begin{tabular}{@{}l@{\hspace{2mm}}
l@{\hspace{3mm}}l@{\hspace{3mm}}l@{\hspace{3mm}}c@{\hspace{3mm}}l@{\hspace{3mm}}
c@{\hspace{3mm}}c@{\hspace{3mm}}c@{\hspace{3mm}}c@{\hspace{3mm}}c
@{\hspace{3mm}}c@{\hspace{3mm}}c@{\hspace{3mm}}c@{\hspace{3mm}}c@{\hspace{3mm}}
c@{\hspace{3mm}}r}
\hline
Star & Study & V & $d$ & $E_{\rm B-V}$ & $T_{\ast}$ &$R_{\ast}$ & log $L_{\ast}$   
& log $\dot{M}$      
& $v_{\infty}$ & H/He&C/He&O/He &log~$Q_{0}$ & log~$Q_{1}$  & $M_{\rm V}$&\\
     &       & mag & kpc & mag & kK         &$R_{\odot}$&$L_{\odot}$& 
$M_{\odot}$~yr$^{-1}$&km\,s$^{-1}$&   &    &     & s$^{-1}$&  s$^{-1}$& mag&\\
\hline
M4--18   &This work  &14.1&6.8&0.55&31  & 2.4 & 3.72 & $-$6.0 & 160 &$<$0.5 & 0.60 &
0.10 &47.2&36.7&$-$1.8\\ 
         &LHJ&13.3&3.5&0.70&31 & 2.3 & 3.67 & $-$6.0 & 350 &$<$10 & 0.40 & 0.05 & &    &$-$1.6\\ 
\noalign{\smallskip}
He~2--113&DC&11.9&1.2&1.00&31 & 2.5 & 3.72 & $-$6.1 & 160 & 0.0 & 0.55 & 0.10
&47.2&36.7&$-$1.8\\ 
\noalign{\smallskip}
\hline
\end{tabular}
\end{center}
\end{table*}

\begin{figure}
\vspace{8.8cm}
\includegraphics{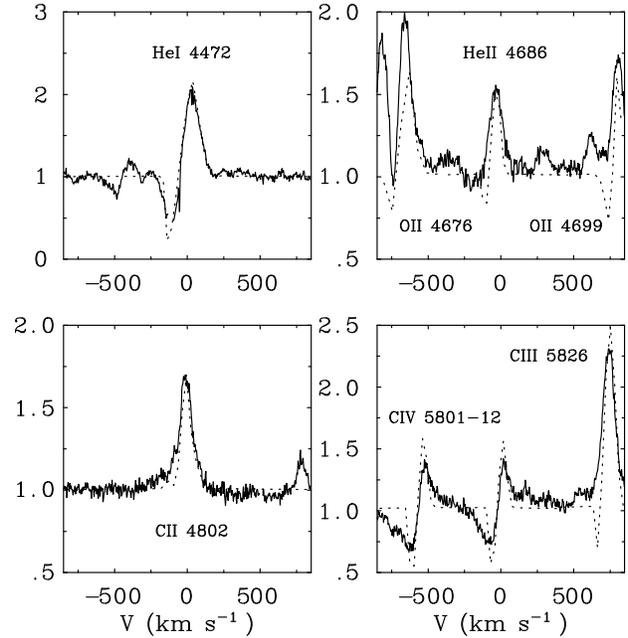}
\caption{A comparison between selected theoretical profiles for M4--18
(dotted lines) with WHT--UES observations (solid lines). Other 
profile fits are comparable to those shown for He\,2--113 in DC.}
\label{fg:fits}
\end{figure}

Fig.~\ref{fg:fits} presents a comparison of selected observed (WHT--UES)
line profiles (solid) with our synthetic spectra (dotted) for M4--18.
Overall, the quality of these and other profile fits,
covering a wide range in excitation and ionization, is comparable with that 
achieved for other [WC10] stars by  DC, where a more comprehensive
explanation of the successes and failures of the model is presented. 
Overall the fits successfully reproduce the strength and shape of many optical
emission line profiles. In particular the helium spectrum is well fitted,
although it is clear from the poor fit to the width of 
He~{\sc ii} $\lambda$4686 (Fig.~\ref{fg:fits} - top right panel) 
that the adopted $\beta$=1 velocity law is not ideal for this star. 
P Cygni absorption components are poorly predicted (e.g. C~{\sc iv} 
$\lambda\lambda$5801--12), further supporting an inadequate velocity 
structure.

Table~\ref{tb:wr_mod} presents a summary of our derived stellar parameters.
Overall, as reflected in their spectra, the stellar properties
and chemistries derived for M4--18 and He~2--113 are almost identical,
except for a marginally stronger stellar wind and higher carbon content
in the former. For reasons discussed by DC
we consider that derived parameters are not critically dependent on our
neglect of heavy element line blanketing. Although we favour a 
non-stellar origin for the pedestal at the base of the H$\beta$ line, we 
have determined a strict upper limit for the stellar hydrogen abundance from 
fitting the wings of the broad H$\beta$ feature, yielding H/He$<$0.5,
by number. Our upper limit, imposes a tighter constraint on the
hydrogen abundance than the analysis of LHJ; this
is solely due to the higher spectral resolution of our 
observations (10 km\,s$^{-1}$ versus $\sim$150 km\,s$^{-1}$).

LHJ included M4--18 in their
quantitative study of [WCL] CSPNe. Although their study used
an independent code, the same assumptions of spherical
symmetry and homogeneity were made. Table~\ref{tb:wr_mod} also compares 
our derived stellar parameters with those from LHJ. 
Overall agreement is very good, although LHJ adopt a 
higher terminal velocity, again due 
to the lower resolution of their data set.
LHJ derive somewhat lower carbon and oxygen contents for M4--18. 

Note that the distance to M4--18  adopted by LHJ
is inconsistent with our value, despite deriving an almost 
identical stellar luminosity. This is because LHJ
obtained an extinction of $E_{\rm B-V}$=0.7\,mag on the basis 
of a spectral fit to {\it exclusively} UV data, inconsistent 
with our stellar and nebular data sets. Their distance estimate
of 3.5\,kpc (following Cudworth 1974)\nocite{C74} 
then led to a visual continuum 
magnitude of $v$=13.3 mag, which our spectrophotometry does not 
support (Section~\ref{sc:mag}).

\section{The M4--18 nebula}\label{sc:hst}

\begin{figure*}
\vspace{3.5in}
\includegraphics{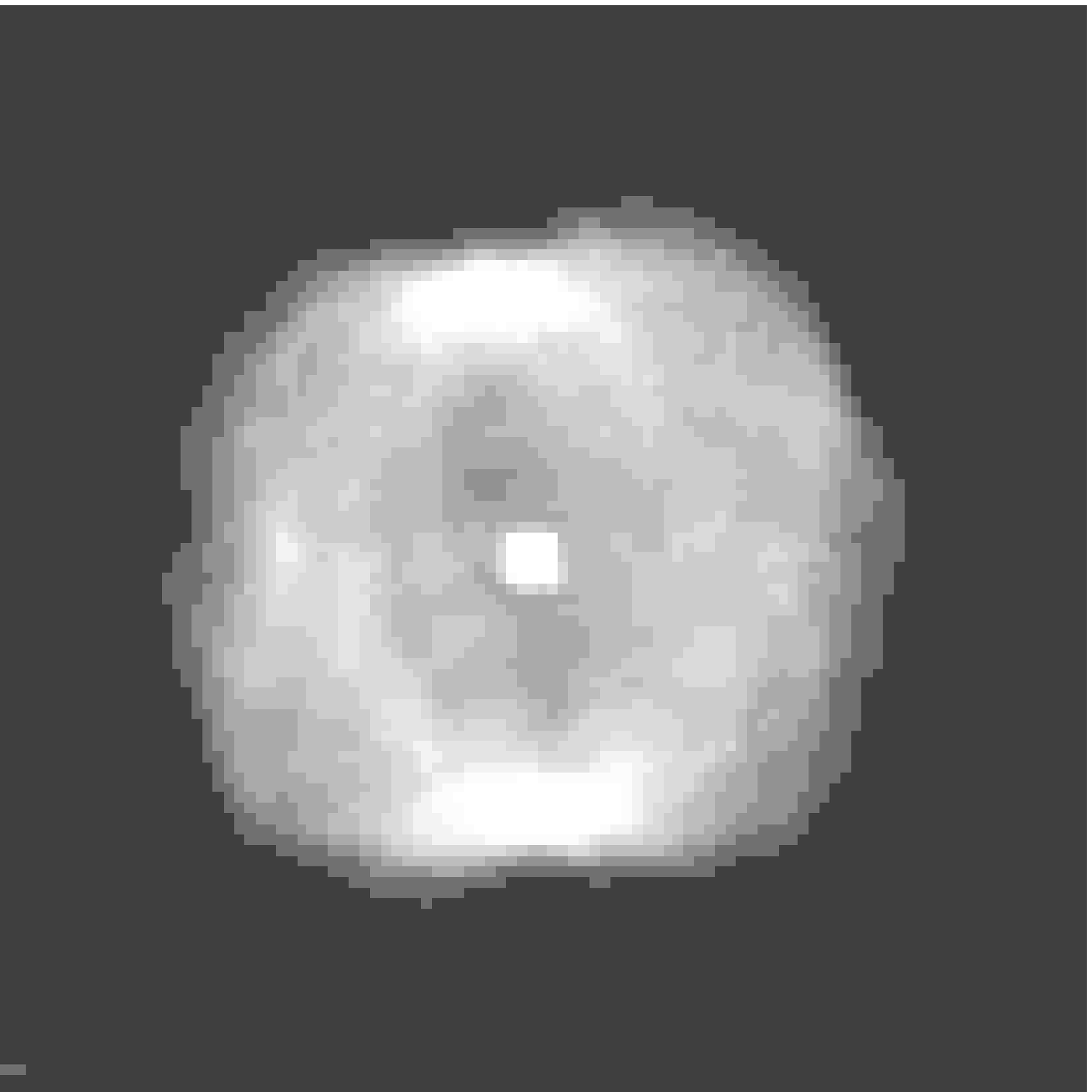}
\includegraphics{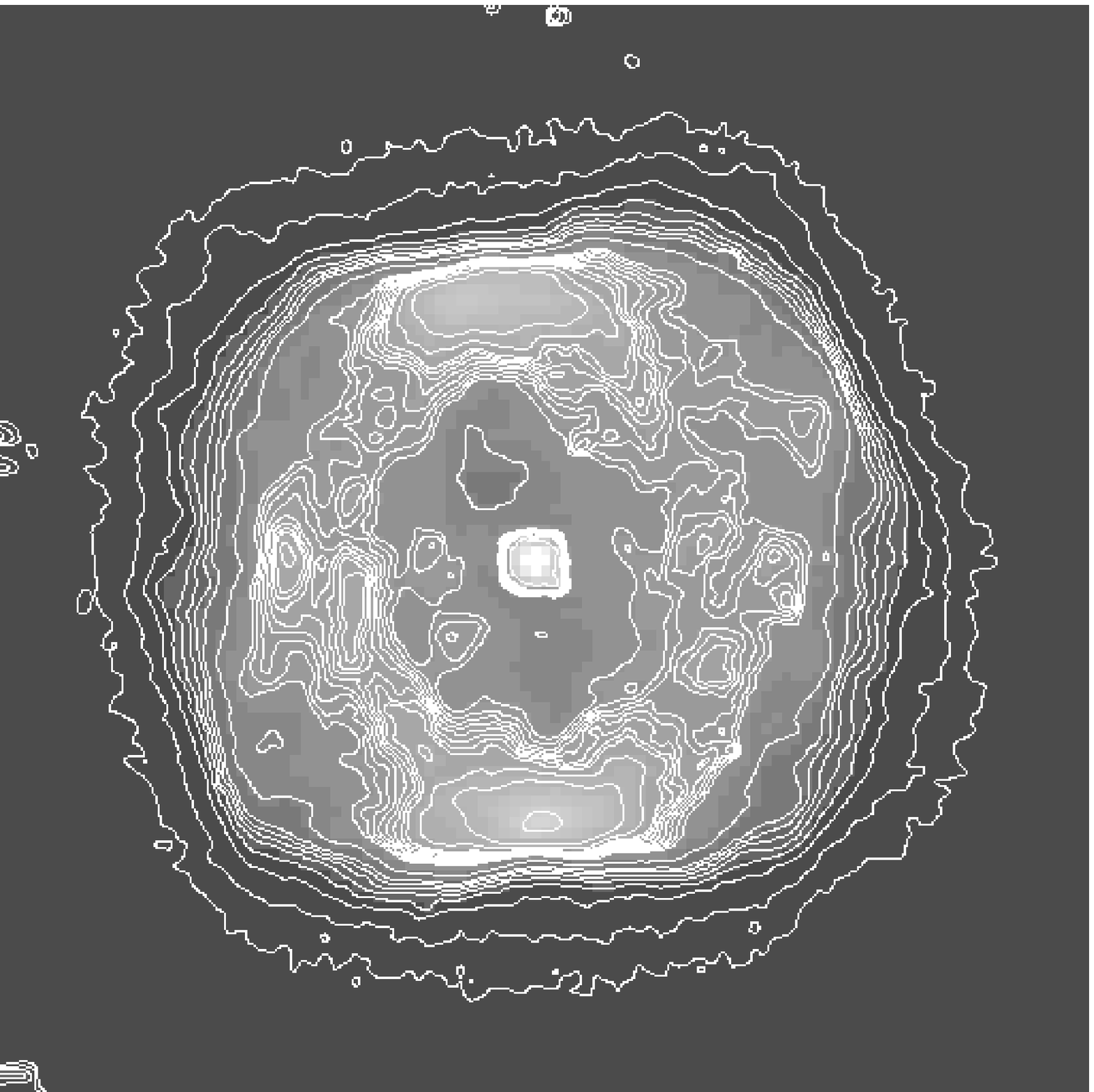}
\caption[HST image of M4--18]{The HST/WFPC2 narrow-band H$\alpha$ (F656N) image 
of M4--18, shown as a grey--scale plot (left) and a contour plot (right). 
We present the combined data sets (u35t2401, u35t2402)
of R.~Sahai, each of 350~sec duration. North is towards the top, 
east to the left while the field of view is 4.5$\times$4.5$''$.}
\label{fg:hst}
\end{figure*}

We now discuss the PN associated with M4--18, including archive
HST/WFPC2 narrow-H$\alpha$ (F656N) images, obtained on 
April 18, 1996 by R.~Sahai. From previous ground based observations, the 
PN of M4--18 is known 
to be extremely small. Radii from optical, mid-IR and radio data sets suggest
$\approx$2$''$ (Shaw 1985; Zijlstra, 
Pottasch \& Bignell 1989)\nocite{S85}\nocite{ZPB89}. 
Although the HST/WFPC2 H$\alpha$ image has recently been presented by
Sahai \& Trauger (1998), we show it in Fig.~\ref{fg:hst} in connection
\nocite{ST98}
with further measurements of its size which are relevant to our analysis.
We will abstain however from discussing its morphology, which is described in
detail by Sahai \& Trauger (1998).
Fig.~\ref{fg:hst} shows the nebula both as 
a grey-scale and a contour map, indicating a clear elongated ring which 
peaks in intensity at a semi-major axis of 1.28$''$ (PA=0)
and a semi-minor axis of 0.8$''$ (PA=90). The ring is embedded in a
larger, less eccentric shell. From an azimuthal average of the PN,
we determine a mean radius of $1.85'' \pm 0.10''$.

At a distance of 6.8 kpc, the radius of M4--18 corresponds to
a physical radius of 0.06 pc (10$^4$ AU). 
Using the observed nebular expansion
velocity we obtain a dynamical age of $\tau_{\rm dyn}$$\sim$3\,000 years. 
Note that this dynamical age should be reasonable since
as we will later show, the PN is optically thin 
(Section~\ref{sc:neb_mod}).

\section{Nebular Abundance Analysis}\label{sc:neb_abu}

Spectroscopy of the M4--18 PN has previously been
carried out by Sabbadin (1980), Goodrich \& Dahari 
(1985)\nocite{GD85}\nocite{S80}
and SR. 
The latter groups combined low resolution optical and 
UV spectroscopy
to determine PN physical conditions and abundances, while
SR also applied their own nebular 
modelling code. We shall now re-derive the electron density 
and temperature for the PN of M4--18, together with elemental
abundances.

\subsection{Nebular line fluxes}
\label{ssc:neb_lin_flu}

Our analysis of the PN associated with M4--18 is 
based on our INT spectrum, in which stellar and nebular features
are blended. (Since the nebular diameter is $\sim$3.7$''$, 
ground-based long-slit spectroscopy do not allow off-star nebular
extraction). We present observed and de-reddened line fluxes in
Table~\ref{tb:neb_flu}.

The [O~{\sc ii}] doublet at 3726 and 3729\AA\ may be contaminated by
a component of the stellar O\,{\sc ii} M2 triplet $\lambda$3727.3.
This was judged to be negligible since the 
un-blended components of this triplet (at 3712.7 and 3749.5\AA)
were weak. However, this doublet was located close to the end of 
our spectral range, where the sensitivity of the
instrument drops sharply.
We therefore assign a 30\% error to our measurements.
Inspection of our UES data set reveals that both 
the [O{\sc iii}] lines at 5007 and 4959\AA\ are present\footnote{The 
[O{\sc iii}] line at 4959\AA\ is heavily blended with C~{\sc ii} stellar 
components (i.e. the four components of the dielectronic Multiplet 25), 
but as we have fitted this spectral region in the spectra of 
CPD--56$^{\circ}$8032 and He~2--113 which did not have any [O~{\sc iii}]
(De Marco, Storey \& Barlow 1998) 
it was straightforward to recognise its presence.}, though extremely weak.

Following the approach discussed in De Marco et al. (1997)
for the C~{\sc iii}] $\lambda$1909 line, we consider this to be entirely
of stellar origin. Instead, we rely on C~{\sc ii}] $\lambda$2326 for 
the determination of the nebular carbon abundance. Its observed equivalent 
width is 15$\pm$4\AA\ from our dereddened IUE spectrum, of which
$\sim$4\AA\ is of stellar origin according to our stellar analysis 
(Section~\ref{stellar}). Since the continuum flux at
2326\AA\ is in good agreement with our reddened theoretical model
distribution (Fig.~\ref{fg:ste_flu}), we are able to obtain
a nebular $\lambda$2326 flux of 5.34$\times$10$^{-12}$ 
erg~cm$^{-2}$~s$^{-1}$.

For sulphur, only [S~{\sc ii}] $\lambda$$\lambda$6713,31 were available.
The [S~{\sc iii}] $\lambda$3722 line is extremely weak (and blended with 
[O~{\sc ii}] $\lambda$$\lambda$3726,29), while the 
$\lambda$$\lambda$9069,9532 lines were outside our observed spectral 
range. SR derived the abundance of
S$^{2+}$/H$^+$, although they derived fluxes for $\lambda$$\lambda$9069,9532.
from indirect techniques.
Finally, for nitrogen we used lines at 5755 and 6548\AA, excluding
[N~{\sc ii}] $\lambda$6584 which is blended with the stellar 
C~{\sc ii} line at 6578\AA.

\begin{table}
\begin{center}
\caption{Observed (F) and dereddened (I; $E_{B-V}$=0.55\,mag) nebular 
line intensities for M4--18 from our INT/IDS spectrum ($\ast$: WHT/UES). 
The  dereddened H$\beta$
flux is 3.62$\times$10$^{-12}$ erg~cm$^{-2}$~s$^{-1}$. Errors are indicated.}
\label{tb:neb_flu}
\begin{tabular}{@{}lllll@{}}
\hline
Ion & $\lambda$ & F                        & 100$\times$I &Errors  \\
    &  (\AA)    &(ergs~cm$^{-2}$~s$^{-1}$) &/I(H$\beta$)  &\%  \\
\hline
$[$O~{\sc ii}] & 3726.0 & 5.69$\times$10$^{-13}$& 158& 30\\
$[$O~{\sc ii}] & 3728.8 & 3.01$\times$10$^{-13}$& 83.7 & 30\\
H$\beta$       & 4861.3 & 5.75$\times$10$^{-13}$& 100  & 10\\
$[$O~{\sc iii}]& 5006.8$^{\ast}$ & $\le$3.6$\times$10$^{-15}$& 0.06 & 50  \\
$[$N~{\sc ii}] & 5754.6 & 1.11$\times$10$^{-14}$& 1.35  & 10\\
$[$O~{\sc i}]  & 6300.3 & 2.15$\times$10$^{-14}$& 2.2  & 10\\
$[$O~{\sc i}]  & 6363.8 & 9.73$\times$10$^{-15}$& 1.0  & 10\\
$[$N~{\sc ii}] & 6548.0 & 4.44$\times$10$^{-13}$& 43.1 & 10\\
H$\alpha$      & 6562.8 & 2.69$\times$10$^{-12}$& 260  & 10\\
$[$N~{\sc ii}] & 6583.4 & 1.54$\times$10$^{-12}$&  147 &20\\ 
$[$S~{\sc ii}] & 6716.5 & 7.19$\times$10$^{-14}$& 6.7  & 10\\
$[$S~{\sc ii}] & 6730.8 & 1.41$\times$10$^{-13}$& 13.0 & 20\\
\hline
\end{tabular}
\end{center}
\end{table}

\subsection{Nebular temperatures, densities and abundances}
\label{ssc:neb_tem_den}

\begin{figure}
\vspace {7.25cm}
\includegraphics{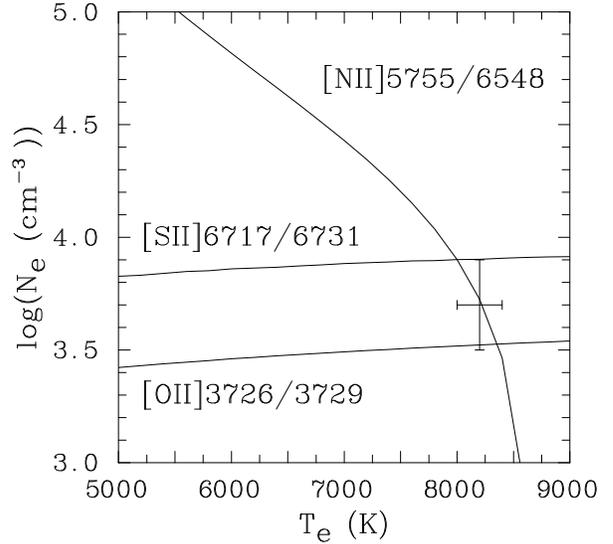}
\caption{Nebular Diagnostic diagram for M4--18.}
\label{fg:ne_te}
\end{figure}
We can now proceed to obtaining estimates of the nebular conditions
and chemistry. {}From the usual
diagnostic diagram relating H$\alpha$/[S\,{\sc ii}] to [S\,{\sc ii}] 
$\lambda$6717/$\lambda$6731 (see Sabbadin, Minello \& Bianchini 1977)\nocite{SMB77} we
find that M4--18 falls in the photo-ionization dominated region, 
rather than being ionized by shock waves. Therefore the usual 
nebular diagnostic techniques may be applied here. 

We have constructed a nebular diagnostic diagram for M4--18 
using [N\,{\sc ii}], [S\,{\sc ii}] and [O\,{\sc ii}] line ratios.
These are generated using the {\sc ratio}
program, written by I.D.~Howarth and S.~Adams which solves the
equations of statistical equilibrium allowing a determination of $N_{e}$
as a function of $T_{e}$ for each ratio. 
{}From Fig.~\ref{fg:ne_te} we determine $N_e$ =10$^{3.7\pm0.2}$~cm$^{-3}$
and $T_e$=8\,200$\pm$200~K for M4--18. The electron density is adopted 
as a compromise between the [O~{\sc ii}] and [S\,{\sc ii}] doublets.
Our results are in reasonable agreement with Sabbadin (1980) and
SR , though Goodrich \& Dahari (1985) derived 
a somewhat lower electron temperature ($T_e$=5600$\pm$1600 K,
$N_e$=10$^{4.0\pm0.1}$~cm$^{-3}$).

\begin{table}
\begin{center}
\caption{Nebular abundances for M4--18 (on the usual scale where 
$\log$N(H)=12.0). The C/O number ratio is on a linear scale.
The [Ne\,{\sc ii}] line flux is taken from Aitken \& Roche (1982).}
\begin{tabular}{lll}
\hline
Ratio &  $\log$(X/H)+12.0 & Diagnostic Lines \\
\hline
C&  9.08$\pm$0.03 & C~{\sc ii}] $\lambda$2326\\
N&  7.60$\pm$0.04 & [N~{\sc ii}] $\lambda$5755,$\lambda$6548\\
O&  8.62$\pm$0.01 & [O~{\sc ii}] $\lambda\lambda$3726,29\\
Ne& 6.73$\pm$0.05 & [Ne~{\sc ii}] 12.8 $\mu$m \\
S&  6.16$\pm$0.06 & [S~{\sc ii}] $\lambda\lambda$6713,31\\
C/O&  2.9$\pm$0.6 \\
\hline
\end{tabular}
\label{tb:abu}
\end{center}
\end{table}

We now utilise our derived nebular parameters to obtain estimates
of elemental abundances, which are listed in Table~\ref{tb:abu}.
For oxygen, nitrogen and carbon we believe the abundances derived from the 
singly ionized ions to be representative of the whole ionized populations since 
[O~{\sc iii}] $\lambda$5007 and C~{\sc iii}] 
$\lambda$1909 are extremely weak or absent.
In the case of oxygen, from the upper limit of $\lambda$5007, we 
derive O$^{2+}$/H$^{+}$$<$4.4$\times$10$^{-7}$, corresponding to 
O$^{2+}$/O$^{+}$$<$1.0$\times$10$^{-3}$.

Equally for sulphur, S$^+$/H$^+$ should be representative of S/H since
SR obtained S$^{2+}$/S$^+$$\sim$0.1. 
Our C/O ratio is much lower than found by De Marco et al. (1997)
for CPD--56$^{\circ}$8032 and He~2--113, which suggests that the
carbon abundance in the PNe of WC central stars is not always 
higher than average.

Overall,
agreement with the abundances obtained by SR is 
good, except for oxygen (they derived log(O/H)+12=8.1) since their 
de-reddened [O\,{\sc ii}] $\lambda$$\lambda$3727,29 flux was almost 10 times
lower than our measurement. Abundances derived by Goodrich \& Dahari (1985) 
are also in reasonable agreement once the different reddening and electron
properties are taken into account.

\section{Nebular Modelling}
\label{sc:neb_mod}

In this Section we derive Zanstra temperatures for M4--18
using blackbody and Kurucz (1991) model atmosphere flux distributions, which are
compared to WR flux distribution resulting from the analysis of the 
CSPN (Section~\ref{stellar}). The nebular radial density profile
is derived from the HST H$\alpha$ image surface brightness 
distribution following the
formalism of Harrington \& Feibelman (1983). We will finally
carry out photo-ionization modelling of the PN
of M4--18 using all available atmosphere flux distributions
and we will finally compare the modeled parameters with those derived
empirically in Section~\ref{sc:neb_mod}.

\subsection{Zanstra temperatures}
\label{ssc:zan_tem}

We derive a flux of hydrogen--ionizing photons corresponding 
to log $Q_0$=46.81 s$^{-1}$ from our H$\beta$ flux (Table~\ref{tb:hflux}),
de-reddened by $E_{\rm B-V}$=0.55 at a distance of 6.8 kpc. 
We derive a blackbody H~{\sc i} Zanstra temperature of 29\,500$\pm$500~K,  
while Kurucz ATLAS9 (Kurucz 1991)\nocite{K91} model atmospheres 
yield an effective temperature of 33\,000$\pm$500~K, using 
log~$g$=4.0 models (the minimum gravity tabulated at the required 
effective temperature). The inferred bolometric luminosities are
4000~$L_\odot$ and 5600~$L_\odot$, respectively. For comparison,
Goodrich \& Dahari (1985) derived a temperature of 22\,000~K 
using the Stoy (1933)\nocite{S33} energy--balance method. 
SR derived a blackbody H~{\sc i} Zanstra temperature of 26\,000K. 
(They also fitted a blackbody to the optical and UV energy distribution 
implying 23\,000~K.)

Our WR stellar model for M4--18 predicts log $Q_0$=47.23 s$^{-1}$.
This is 2.6 times greater than that inferred from the nebular H$\beta$ 
flux. In contrast, DC showed how the WR model 
atmosphere ionizing flux of He~2--113 was 85 times larger than that 
predicted by H$\beta$, suggesting that for high density PN, gas--dust 
competition might play a role. Our results for M4--18 confirm 
their suspicion.

\subsection{Nebular surface brightness and density
distributions}\label{sect8.2}

We have obtained the radial density distribution for
M4--18, necessary for photo-ionization modelling, from the 
archival HST/WFPC2 image (Section~\ref{sc:hst}). For this PN 
the contribution from nebular continuum emission is negligible. 

\begin{figure}
\vspace{7.75cm}
\includegraphics{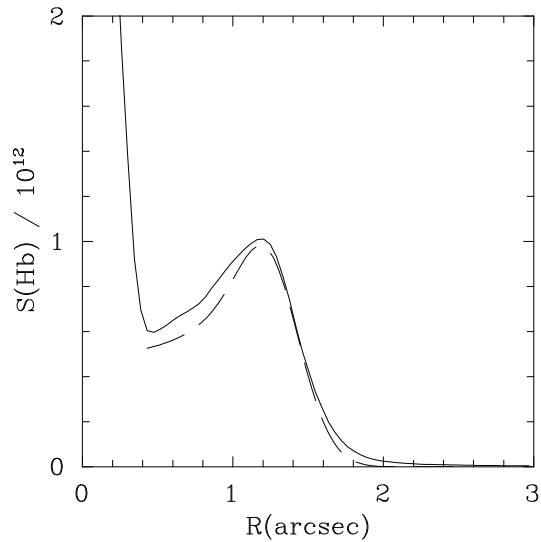}
\caption{H$\beta$ surface brightness profile (in 
units of erg~cm$^{-2}$~s$^{-1}$~arcsec$^{-2}$) derived from the HST 
H$\alpha$ image of M4--18 from an azimuthal average centred on the star 
(solid line; the central star's continuum is responsible for the
intensity peaking near the centre of the nebula). The dashed line
is the intrinsic surface brightness distribution predicted by our
photo-ionization model.}
\label{fg:den_prof}
\end{figure}

The azimuthally averaged surface brightness distribution in the 
H$\alpha$ image was derived using an algorithm implementation in
the {\sc surfphot} package of {\sc midas}. The centre
of the image was chosen to coincide with the position of the central star. 
The integrated flux within a radius $r$ was then determined.
The central star was contained
within a radius of 5 pixels and the contribution of its continuum
flux was subtracted from the nebular flux. The nebular flux, integrated
over the entire PN, was then normalised to the de-reddened H$\beta$
flux, differentiated with respect to $r$ and divided by 2$\pi r$
to obtain the required azimuthally averaged H$\beta$
surface brightness distribution.
The result is plotted in Fig.~\ref{fg:den_prof} (solid line).

For optically thin PNe, the formalism of 
Harrington \& Feibelman (1983) allows
one to derive a density distribution from a surface brightness distribution
(see Liu et al. 1995\nocite{LBBOC95} 
for an implementation of this method). Photo-ionization
models based on density distributions derived with this formalism, however,
tend to give an average density which is too low to reproduce 
density-sensitive line ratios such as the [O~{\sc ii}]
$\lambda$3726/$\lambda$3729 ratio. This deficiency is overcome by introducing
a filling factor, $\epsilon$, such that within any small region of material
a fraction $\epsilon$ is filled with gas, while the complementary
fraction (1--$\epsilon$) is empty. In our case all density
sensitive line ratios were well reproduced with a filling factor
of unity (i.e. a completely filled nebula).

\begin{figure}
\vspace{11cm}
\includegraphics{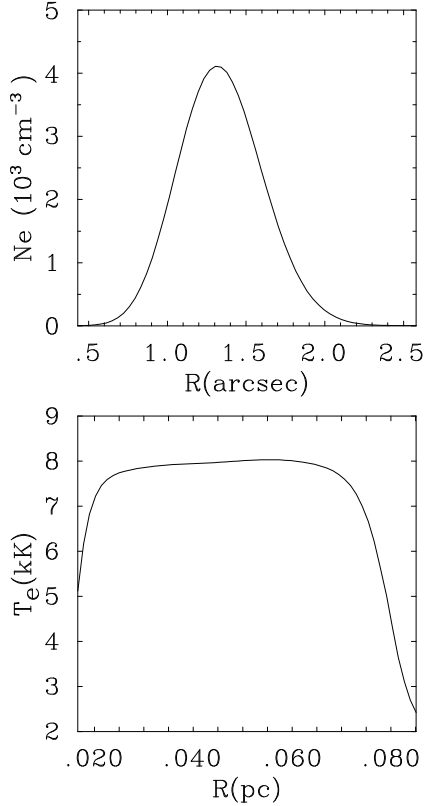}
\caption{(Upper panel): Electron density distribution for 
M4--18 obtained from our analysis of the HST/WFPC2 image; 
(Lower panel): Electron temperature resulting from our 
photo-ionization modelling with the WR flux distribution
(assuming a distance of 6.8kpc).}
\label{fg:mod_te_ne}
\end{figure}

The density distribution which reproduces the H$\beta$ surface brightness
distribution, absolute H$\beta$ value and density sensitive line ratios
is shown in Fig.~\ref{fg:mod_te_ne} (upper panel). 
To reproduce the surface brightness shown in Fig.~\ref{fg:den_prof}
we used $N=12$ and $K=1$ in equation A2 of Harrington \& Feibelman (1983).\nocite{HF83}
The upper panel of Fig.\ref{fg:mod_te_ne} shows that M4--18 has a 
central cavity surrounded by a thick shell whose density declines fairly 
symmetrically as the radius increases.

\subsection{The photo-ionization modelling}

The photo-ionization code used to model M4--18 
(Harrington et al. 1982) assumes that the nebula can be represented
by a hollow, spherical shell which is ionized and heated solely by
the radiation of the central star. This shell is sampled by a
network of grid points at each of which the density is supplied. 
We used 60 grid points and the electron density derived from the 
H$\beta$ surface brightness (Section~\ref{sect8.2}).

Once a distance is adopted, and the de-reddened V magnitude
of the central star established, the effective temperature
of the central star (from the WR stellar analyis or derived by
a Zanstra analysis as described in Section~\ref{ssc:zan_tem}) determines
all the other stellar parameters. These parameters are listed
in the first three rows of Table~\ref{tb:neb_mod}.
The inner nebular radius was fixed at the value derived from 
the HST/WFPC2 observations, scaled to the adopted distance. 
Elemental abundances are initially set at empirical values,
and later adjusted to fit the observed line fluxes if necessary.

Final photo-ionization models for each flux distribution 
are compared to observation in Table~\ref{tb:neb_mod}. Overall,
very good agreement is reached 
between observations and model calculations for all diagnostic line
ratios (O\,{\sc ii}, N\,{\sc ii} and S\,{\sc ii}) as well as 
absolute line fluxes. This is an indication of the success of our density 
distribution, recalling the discrepancy between different diagnostics in 
Fig.~\ref{fg:ne_te}. 

It is apparent that the WR and 
Kurucz (1991) stellar flux distributions perform better than the blackbody.
Overall, the WR stellar atmosphere is most appropriate to 
reproduce the line intensities, in spite of the other distributions 
resulting from Zanstra nebular analyses. In Fig.~\ref{fg:mod_te_ne}, we
present the predicted electron temperature as a function of radius 
for our WR model. Hydrogen is completely ionized throughout,
as is appropriate for an optically thin PN, with a predicted
nebular mass of 0.08$M_{\odot}$.

\begin{figure}
\vspace{8.75cm}
\includegraphics{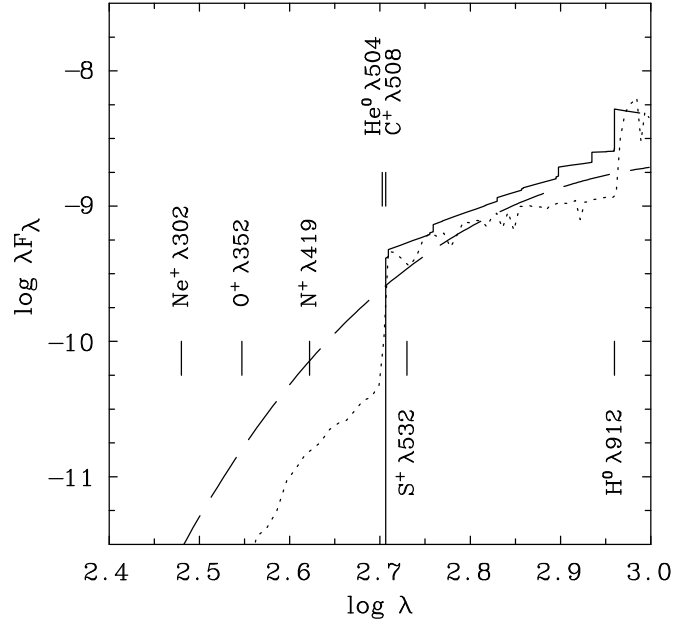}
\caption{Comparison of Lyman ionizing flux distribution
obtained from our stellar analysis of M4--18 (Wolf-Rayet, solid), 
with black-body (dashed) and Kurucz (1991, dotted) distributions obtained
from our nebular Zanstra analysis. Clearly the WR distribution differs
dramatically from the other distributions shortward of the He\,{\sc i} edge
at $\lambda$504, especially at the O\,{\sc ii} edge at $\lambda$353.}
\label{zanstra}
\end{figure}

A major failure of each model flux distribution is the 
overestimate of the sulphur ionization balance, and consequent
underestimate of the [S~{\sc ii}] 6717,31\AA\ line strengths.
(The ratio S$^{2+}$/S$^{+}$ was estimated to be
0.1 by SR, although a value of $<$1 
is probably more secure.)
The poor reproduction of the sulfur ionization balance can be appreciated
by inspecting the shape of the ionizing flux distributions in
Fig.~\ref{zanstra}. Metal line blanketing in our WR model atmospheres 
may help to solve the sulphur problem by reducing the hard UV flux
(for the effect of blanketing in late-type WR stars see 
Crowther, Bohannan \& Pasquali 1998)\nocite{CBP98}. 

In the case of the oxygen ionization balance, only
the WR model predicts the appropriate ratio, with Kurucz (1991) and blackbody 
atmospheres overestimating the strength of [O~{\sc iii}] $\lambda$5007.
This is illustrated
in Fig.~\ref{zanstra} where only the WR energy distribution
has negligible zero flux below the He\,{\sc i} edge at $\lambda$504.

\begin{table}
\begin{center}
\caption
{Comparison of results from photo-ionization models for
M4--18 with observations (I(H$\beta$)=100),
using WR non--LTE, blackbody and Kurucz (1991) plain--parallel energy 
distributions. Elemental abundances are by number. 
Filling factors of $\epsilon$=1 are assumed throughout.
}
\label{tb:neb_mod}
\begin{tabular}{l@{\hspace{-1.5mm}}c@{\hspace{1mm}}c@{\hspace{3mm}}c@{\hspace{3mm}}c}
\hline 
Parameter & Empirical &  WR & Blackbody & Kurucz \\
\hline 
$T_{\em eff}$(kK)    & --   & 31.0     & 29.5  & 33.0  \\
log($L_{\ast}/L_{\odot})$   & --   & 3.72     & 3.60  & 3.75  \\        
log $Q_{0}$(s$^{-1}$)&46.81 & 47.23    & 46.81  & 46.81  \\    
$N_{e}(r)$(cm$^{-3}$)&\multicolumn{4}{c}{See Figure~\ref{fg:mod_te_ne} (upper panel)}  \\
\noalign{\smallskip}
He/H                   &--   & 0.1 & 0.1 & 0.1 \\
C/H$\times$10$^3$      & 1.2 & 1.2 & 1.2 & 1.2 \\
N/H$\times$10$^5$      & 4.0 & 4.8 & 5.5 & 4.0 \\
O/H$\times$10$^4$      & 4.2 & 4.2 & 4.2 & 4.2 \\
Ne/H$\times$10$^5$     & 5.0 & 4.1 & 4.1 & 4.0 \\
S/H$\times$10$^6$      & 1.4 & 1.4 & 1.4 & 1.4 \\
\noalign{\smallskip}
$[$O~{\sc ii}]$\lambda$3726/$\lambda$3729 &1.88 &1.82 &1.85 & 1.80 \\
$[$S~{\sc ii}]$\lambda$6716/$\lambda$6731 &0.52 &0.60 &0.60 & 0.62 \\
$[$N~{\sc ii}]$\lambda$5755/$\lambda$6548 &0.032&0.027&0.029& 0.032 \\
\noalign{\smallskip}
   C\,{\sc iii}]   $\lambda$1909& 0   & 0.0  & 11    & 2.6 \\
  C\,{\sc ii}]     $\lambda$2326& 148 & 128  & 96    & 197 \\
 $[$O\,{\sc ii}]     $\lambda$3726& 158 & 155  & 154   & 213  \\
 $[$O\,{\sc ii}]     $\lambda$3729& 84  & 85   & 83    & 118  \\
 $[$O\,{\sc iii}]    $\lambda$5007& 0   & 0.0  & 39    & 1.0\\
 $[$N\,{\sc ii}]     $\lambda$5755& 1.4 & 1.13 & 1.14  & 1.35\\
 $[$O~{\sc i}]       $\lambda$6300& 2.2 & 0.049& 0.067 & 0.23 \\
 $[$O~{\sc i}]       $\lambda$6363& 1.0 & 0.016& 0.022 & 0.075\\
 $[$N\,{\sc ii}]     $\lambda$6548& 43  & 42   & 41    & 42  \\
 $[$S\,{\sc ii}]     $\lambda$6717& 6.7 & 0.64 & 0.87  & 1.29\\
 $[$S\,{\sc ii}]     $\lambda$6731& 13  & 1.73 & 1.85  & 2.08\\
$[$Ne\,{\sc ii}] 12.8$\mu$m & 32  & 32   & 32    & 32  \\
\noalign{\smallskip}
S$^{2+}$/S$^{+}$       & $<$1&12    & 15    & 17  \\
O$^{2+}$/O$^{+}$       &$<$1.0$\times$10$^{-3}$&$\sim$0 &77$\times$10$^{-3}$&1.6$\times$10$^{-3}$\\
$T_{e}$(N$^+$)(K)      &8200 &7900  & 8000  &8400\\
$\tau$(H~{\sc i})    &--    & 1.1      & 1.6     & 4.2 \\  
\hline
\end{tabular}
\end{center}
\end{table}

Our study supersedes the photo-ionization modelling of M4--18 by 
SR. They adopted a distance of 0.9kpc, and
a 22\,000~K blackbody, implying a luminosity of 134$L_\odot$ for the 
central star. As discussed in Section~\ref{distance}, these quantities are 
inconsistent with theoretical predictions for CSPNe. Note also that 
SR also adopted a constant electron density of 6620~cm$^{-3}$, implying 
$T_{e}$=6\,600~K (their empirical value was 8\,500~K). 

\section{Discussion}\label{sc:disc}

We have derived consistent stellar and
nebular parameters for the [WC10] central star M4--18 and its PN.
Although stellar and nebular analyses
have previously been carried out (LHJ; SR) such studies revealed inconsistent parameters
(e.g. LHJ disregarded its visual magnitude
while SR adopted an unrealistically low luminosity). 

While the stellar parameters of M4--18 are almost 
identical to those of He~2--113, their PNe differ greatly.
For M4--18, a dynamical age of 3\,100 years
is implied for a distance of 6.8~Kpc, while the well determined distance
to He~2--113 (De Marco
et al. 1997), implies a dynamical age for this PN of $\approx$270~yr.
Since He~2--113 is optically thick, its dynamical age is probably
only a lower limit, although there can be little doubt that the PN
of He~2--113 is younger, when we consider the respective
optical depths and electron densities. How can such spectroscopically
similar CSPN have such different PN ages? This difference is hard to reconcile,
if M4--18 and He~2--113 have similar masses. 

Were we to relax our assumption that all [WC]-type stars have 
an identical mass, distances in the range 4.2--12~kpc are 
consistent with our  spectrophotometry and the luminosity from 
theoretical He-burning tracks. Table~\ref{tb:disc} 
shows the properties of M4--18 for this range of distances
compared to those of He~2--113 (De Marco et al. 1997;
DC). We include
the ages predicted by the 
evolutionary tracks of Bl\"ocker (1995) and Vassiliadis \& Wood (1994).
(Note that these agree well where comparisons may be carried out).
Ages are predicted to be extremely sensitive to core mass. From 
Figures 15 
and 16 of Bl\"ocker (1995), a helium-burning central star of mass 
0.52~M$_\odot$ reaches 30\,000~K in 8\,000 years, while 
a star of 0.84~M$_{\odot}$ takes only 150 years.

Comparisons between
dynamical and evolutionary ages should be treated qualitatively
(both dynamical and evolutionary ages of the PNe are subject to large
uncertainties). Nevertheless, assuming a lower mass for M4--18 
than He~2--113 resolves the differences in their PNe.
In addition, it resolves the discrepancy between the dynamical 
and evolutionary ages of M4--18. For a mass of $\sim$0.55\,M$_\odot$, 
the dynamical and evolutionary time-scales are both $\sim$2500 yr
(corresponding to a distance of 5.5~kpc\footnote{At a distance of 
5.5\,kpc, the luminosity, radius and mass-loss rate for M4--18 would be 
revised to 
log~($L_{\ast}/L_{\odot}$)$\sim$3.45, $R_{\ast}$=1.8$R_{\odot}$ and 
log($\dot{M}$/$M_\odot$~yr$^{-1}$)=--6.25, with other stellar 
properties unchanged.}).
We note that for He~2--113 the comparison between dynamical and 
evolutionary time-scales is reasonable if we take into consideration that the 
dynamical time-scale is merely a lower limit.

\begin{table}
\caption{Summary of stellar and nebular parameters for M4--18,
compared with those for its spectroscopic twin He~2--113 (De Marco
et al. 1997; DC). Evolutionary
time-scales ($\tau_{\rm evol}$) are from Bl\"{o}cker (1995) and Vassiliadis \&
Wood (1994).}
\label{tb:disc}
\begin{tabular}{c@{\hspace{3mm}}r@{\hspace{2.5mm}}c@{\hspace{2.5mm}}
r@{\hspace{2.5mm}}r@{\hspace{3mm}}r@{\hspace{3mm}}c@{\hspace{3mm}}r}
\hline
 $D$ & $N_e$   &$v_{\rm exp}$    & Radius & $\tau_{\rm dyn}$ 
&  $L_{\ast}$  & $M$ & $\tau_{\rm evol}$\\
kpc&cm$^{-3}$&km~s$^{-1}$&10$^{-3}$pc & yr&$L_\odot$&$M_\odot$&yr\\
\hline
\multicolumn{7}{c}{M4--18}\\
 4.8 & 3300        & 19  &43  & 2200  & 2600 & 0.52 & 8\,000 \\
 6.8 & 3300        & 19  &61  & 3100  & 5250 & 0.62 & 500 \\
 12  & 3300        & 19  &111 & 5600  & 16000& 0.84 & 150 \\
\multicolumn{7}{c}{He~2--113}\\
 1.2 & 63\,000     & 21  &6 &$\ge$270  & 5250 & 0.62 & 500 \\
\hline
\end{tabular}
\end{table}

If the masses of M4--18 and He~2--113 are different, then 
masses of WR central stars may span the whole range of
possible post AGB masses, opening the possibility that WR
central stars follow a variety of evolutionary paths.
It has been suggested (e.g. Rao 1987; SR)
that [WC10] central stars are approaching the AGB in a 
born--again type evolution (Iben et al. 1983)\nocite{IKTR83}. This
claim was based on the apparent inverse correlation between PNe 
diameters and CSPN temperatures. Incorporating the results from
DC, this is no longer established. 
Consequently, there is no evidence that [WC10] CSPN are 
approaching the AGB. However, our comparison provides evidence that 
WC central stars are not a homogeneous group, with a range of 
masses and individual evolutionary patterns. 
It remains to be confirmed whether WC-type stars can result from a 
`born--again' evolution, although the potentially related stars
Abell~30 and Abell~78 are associated with well established 
born--again PNe (Jacoby \& Ford 1983)\nocite{JF83}.

\section{Acknowledgments}
We would like to thank John Hillier for kindly providing us with
his atmospheric code and greatly appreciate discussions with
Mike Barlow. John Deacon is thanked for providing filter profiles and 
calibrations. Dave Sawyer
of the National Optical Astronomy Observatories
is gratefully acknowledged for providing the WIYN/DensPak
spectrum of the Na~{\sc i} D lines. OD acknowledges financial support 
from the Perren
Fund during the period of her Ph.D. and the Swiss Research Council jointly 
with  the Institute of Astronomy at the ETHZ--Z\"urich
for the time and resources used in the final stages of this research. 
PAC gratefully acknowledges financial support from PPARC and the Royal
Society. 

The WHT and INT are operated on the island of
La Palma by the Isaac Newton Group in the Spanish Observatorio
del Roque de los Muchachos of the Instituto de Astrofisica de Canarias.
We particularly thank Don~Pollacco for obtaining the WHT/UES spectrum
as part of a service programme.
This research has made use of observations made with the NASA/ESA 
Hubble Space Telescope, obtained from the data archive at the Space 
Telescope Science Institute. STScI is operated by the Association 
of Universities for Research in Astronomy, Inc. under the NASA contract 
NAS5-26555. IRAF was written and supported by NOAO. 
MIDAS is developed and distributed by the European Southern Observatory.
Calculations have been performed at the CRAY J90 of the RAL Atlas centre 
and at the UCL node of the U.K.~STARLINK facility.

\label{lastpage}
\end{document}